\documentclass[a4paper,11pt]{article}
\usepackage{pos}

\usepackage{svg}
\usepackage{gensymb}
\usepackage{wrapfig}
\usepackage{caption}
\usepackage{subcaption}

\graphicspath{{Figures}}

\newcommand{\eV}{\; \text{eV}}

\newcommand{\Xmax}{ X_{\text{max}}}
\newcommand{\gcm}{\; \text{g}/\text{cm}^2}

\newcommand{\textMHz}{$\text{MHz}$}

\newcommand{\textXmax}{$X_{\text{max}}$}
\newcommand{\textgcm}{$\text{g} / \text{cm}^2$}

\newcommand{\slice}{\text{slice}}

\title{Generalising template synthesis of EAS radio emission to other geometries}
 \ShortTitle{Synthesis of EAS radio emission}

\author*[a]{Mitja Desmet}
\author[a]{Stijn Buitink}
\author[a,b]{Tim Huege}

\affiliation[a]{Inter-University Institute For High Energies (IIHE), Vrije Universiteit Brussel (VUB),\\
	Pleinlaan 2, 1050 Brussels, Belgium}

\affiliation[b]{Institut für Astroteilchenphysik (IAP), Karlsruhe Institute of Technology (KIT),\\
	PO Box 3640, 76021 Karlsruhe, Germany}

\emailAdd{mitja.desmet@vub.be}

\abstract{
	Over the last few decades, radio detection has become one of the standard techniques to study high-energy cosmic-ray air showers. For the purpose of analysing the data, we heavily rely on Monte Carlo simulations. Upcoming dense radio array experiments such as LOFAR2.0 and SKA will, however, reach the limit of what is computationally feasible with these. Other techniques are available, based on macroscopic quantities, but their accuracy has thus far not been adequate to use them in precision analyses. In this contribution we present the latest update on the template synthesis approach, a hybrid model using both micro- and macroscopic inputs to synthesise the radio emission for an air shower with an arbitrary longitudinal profile. The method starts from the emission of a given shower and employs semi-analytical relations which only depend on the atmospheric depth at shower maximum and antenna position in order to transform it. Core to the template synthesis approach is the slicing of the atmosphere. By considering the radio emission from each slice separately, we only need to explicitly account for shower age effects. In previous work it was shown this could be done over a wide range of primary energy and across primary types for vertical air showers, with an accuracy of 10\%. Here, we generalise the method to other zenith angles. We investigate the potential to synthesise between different geometries using a data set consisting of several hundreds of CORSIKA showers with primary energies ranging from $10^{17}$ eV to $10^{19}$ eV.
}

\ConferenceLogo{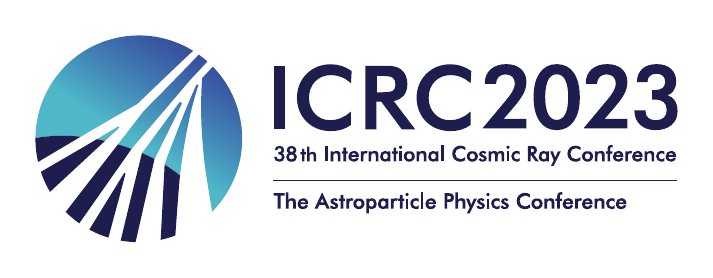}

\FullConference{%
38th International Cosmic Ray Conference (ICRC2023)\\
  26 July - 3 August, 2023\\
  Nagoya, Japan}

\begin{document}
\maketitle

\section{Introduction}

The radio detection of cosmic ray air showers is a technique which has gained renewed interest in the last few decades, with the advent of digital signal processing techniques \cite{Falcke2004, Huege2016}. The low cost associated to the deployment and maintenance of radio antennas makes them an attractive option to instrument large areas, which is necessary to detect cosmic rays at the highest energies where the flux drops to fewer then one particle per $\text{km}^2$ per century.

Instead of measuring these cosmic rays directly, we observe the cascades of particles they induce in the atmosphere. These are referred to as extensive air showers (EAS) and can be detected in several ways. One of them is the radio emission emitted by the charged particles in the cascade, resulting from induced currents and charge imbalances in the cascade \cite{Huege2016}. The first and dominant mechanism in air is the geomagnetic (GEO) emission. It results from the deflection of the charged particles in Earth's magnetic field in combination with the "friction" due to collisions with atmospheric molecules, effectively inducing a transverse current in the shower front. The second effect is often referred to as the charge-excess (CE) component and can be recognised as the Askaryan effect. While propagating through the atmosphere, the shower front accumulates a negative, time-varying charge, which also radiates.

From this radio emission, the properties of the shower and hence those of the primary cosmic ray can be inferred. For example, it has been shown that the atmospheric depth at which the shower has its maximum number of particles, denoted as \textXmax, can be reconstructed from radio data. Furthermore, recent studies show we are also able to reconstruct the width and asymmetry of the particle number longitudinal profile using radio \cite{Mitra2021, Corstanje2022}. All of these parameters inform us about the mass composition of the primary cosmic rays.

However, all of the above mentioned analyses rely on Monte-Carlo (MC) simulations in order to analyse the radio data. One of the current state-of-the-art codes is CORSIKA with the CoREAS plugin for the radio emission. While very precise, the time required to run them poses a stringent limitation on our analyses. This problem is exacerbated by the next generation of cosmic ray observatories, such as LOFAR 2.0 and SKA, which will record data with orders of magnitude more antennas and have much higher event rates. The resulting increase in computation time of the microscopic MC codes will make it unsustainable to use them.

To address this challenge, we have developed template synthesis, a novel method to synthesise the radio emission from cosmic ray air showers. It uses a hybrid approach, employing semi-analytical relations extracted from a set of MC simulations in order to rescale the radio emission between different EAS. Once the relations are obtained for a given air shower geometry, the method only needs a single MC simulation  (the \textit{origin} shower) in order to synthesise the emission from a shower with an arbitrary longitudinal profile (the \textit{target} profile) with the same geometry. From the target profile, template synthesis can obtain the \textXmax\ and shower evolution, both of which are used to transform the emission from the origin into that of the target.

We have applied template synthesis to several geometries individually and obtained good results. That is, for several geometries we generated a dedicated set of MC simulations in order to extract the \textit{spectral functions} for that geometry. In recent work \cite{Desmet2023} we have shown that for vertical air showers we can synthesise the emission with an accuracy of better than 10\% over a broad frequency range of [20, 500] \textMHz. In this contribution, we show the first results for inclined geometries.

\section{The template synthesis approach}

\begin{wrapfigure}[24]{r}{0.55\textwidth}
	\vspace{-10mm}
	\begin{center}
		\includegraphics[width=0.45\textwidth]{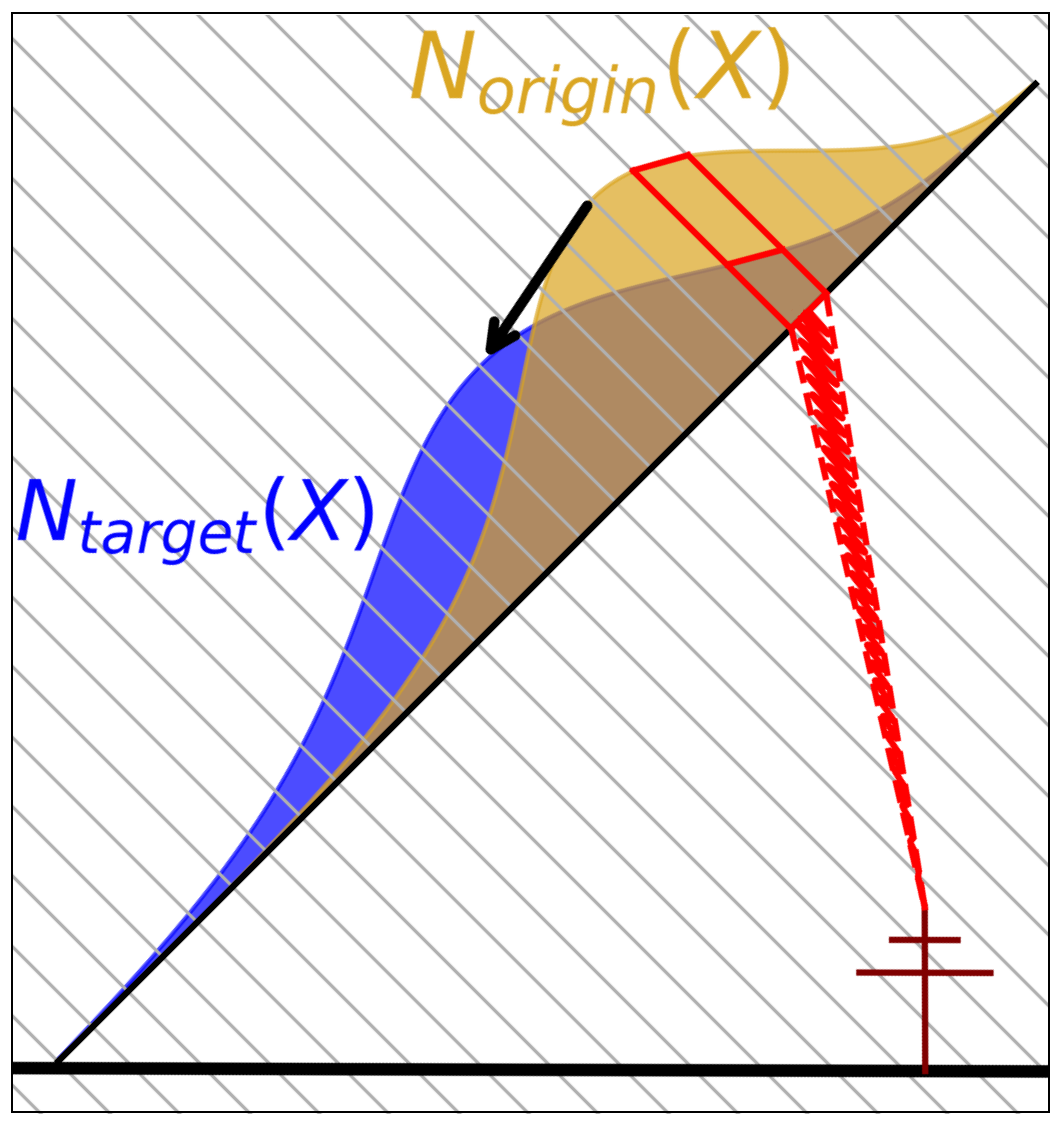}
	\end{center}
	\caption{In template synthesis we divide the atmosphere into slices of constant atmospheric (slant) thickness. The sloping black line represents the depth along the shower axis, with the bold horizontal line being ground. The red lines represent the emission from the selected slice. In our simulations we took the antennas along the $\vec{v} \times \left( \vec{v} \times \vec{B} \right)$ axis. When scaling the emission from $N_{\text{origin}}$ to $N_{\text{target}}$, both the number of particles in the slice as well as the shower age are taken into account.}
	\label{fig:slicing}
\end{wrapfigure}

\begin{figure}
	\begin{subfigure}[t]{0.48\textwidth}
		\centering
		\includegraphics[width=0.95\textwidth]{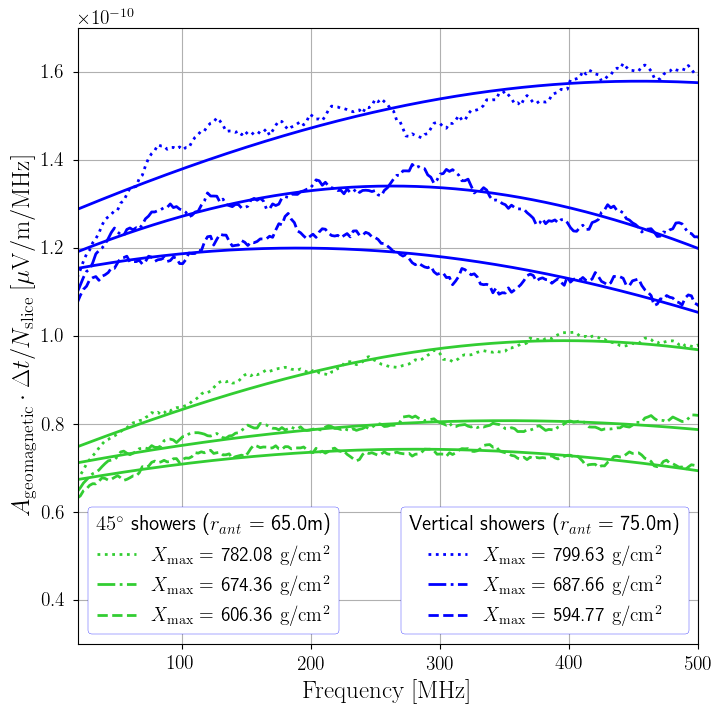}
		\caption{Amplitude spectra for the GEO component of three different showers in a slice at 600 \textgcm, for both the vertical and inclined geometries. The antenna distance in the shower plane is slightly different for each geometry. In both cases we observe a hierarchy according to the \textXmax\ of the shower, namely the peak amplitude increases with \textXmax\ (i.e. decreasing shower age). To each spectrum we fit the function in Equation \eqref{eq:amplitude}, shown by the solid line. The difference in absolute amplitude is mainly due to the larger distance from slice to antenna in the inclined case.}
		\label{fig:fitted_spectra}
	\end{subfigure}
	~
	\begin{subfigure}[t]{0.48\textwidth}
		\centering
		\includegraphics[width=0.95\textwidth]{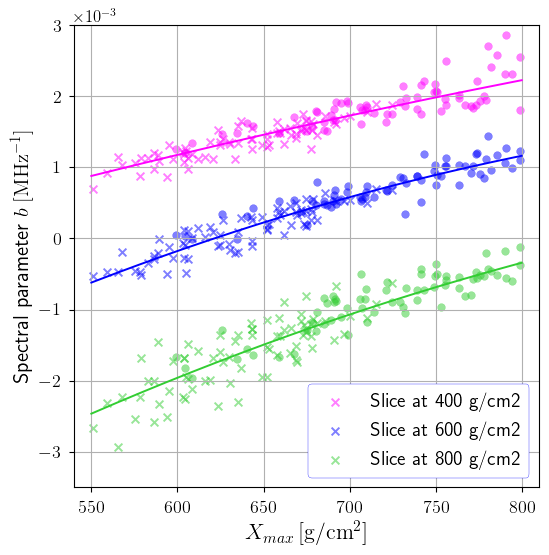}\\
		\caption{The evolution of the spectral $b$ parameter for the GEO component with \textXmax, for a geometry with zenith angle $\theta=45 \degree$. We show three different atmospheric slices, for an antenna at 200m from the axis in the shower plane. The values for proton induced showers are indicated with circles, iron primaries are marked with crosses. The showers also had different primary energies, but the data points still fall on a single parabola. We fit this parabola to obtain the spectral function in each slice, for every antenna separately.}
		\label{fig:spectral_b}
	\end{subfigure}
	\caption{We extract the spectral functions by first fitting a parametrised function to the amplitude frequency spectra in every atmospheric slice (left). When we plot the values of the fitted parameters against the \textXmax\ of the showers they were extracted from, we observe the data points all fall on a single line. We fit a line to the profile of these points, resulting in the spectral functions (right). It is interesting to note that the plots for the vertical geometry look very similar to this one.}
\end{figure}

\noindent
In order to apply template synthesis, we use the idea of sliced simulations as shown in Figure \ref{fig:slicing}. This consists of dividing the atmosphere into slices with a constant atmospheric thickness (in \textgcm) and assigning the radio emission from every particle to the slice where it was emitted\footnote{In CoREAS, which employs the endpoint formalism, this is defined as the slice where the particle starts its track.}. The total signal in an antenna can easily be retrieved by summing those from all slices. Instead of attempting to parametrise the emission from the entire shower in an antenna, we consider the emission from every slice separately. We find its amplitude scales to first order with the number of emitting particles in that slice. On top of this obvious scaling, we also find the pulse shape to depend on the age of the shower in the slice, as shown in Figure \ref{fig:fitted_spectra}.

The concept of shower age is often used in the context of air shower universality \cite{Lipari2009}. It indicates the stage of evolution the shower is in, and is often parametrised as the relative distance to \textXmax. Turning this argument around, within a given slice we can take the \textXmax\ of a shower to act as a proxy for its age. Therefore, within every atmospheric slice we characterise the radio emission in terms of the \textXmax\ to encode the dependency of the pulse shape on the shower age.

Before parametrising the emission from a slice, we separate it into the GEO and CE components. We achieve this by exploiting the fact they are polarised differently \cite{Glaser2016}. In our simulations, we took our antennas on the $\vec{v} \times \left( \vec{v} \times \vec{B} \right)$ axis, where the polarisations of the two mechanisms are orthogonal. For each component we calculate the amplitude frequency spectrum,
\begin{align} \label{eq:dft}
	A_{\text{slice}}(f) = \biggl\lvert \sum_{n=0}^{N} s(n \cdot \Delta t) &\exp( -i 2 \pi \cdot n \cdot f \cdot \Delta t) \biggr\rvert \; ,
	\qquad f \in \left[ 0, \frac{1}{T}, \cdots, \frac{1}{2 \Delta t} \right] \;,
\end{align}
and parametrise each of them with the function
\begin{align} \label{eq:amplitude}
	\tilde{A}_{\text{slice}} (f,  X_{\slice}) = \left( a \cdot N_{\slice} \right) 
	\cdot \exp \left( b \cdot (f - f_0) + c \cdot (f - f_0)^2 \right) 
	\; ,
\end{align}
as shown in Figure \ref{fig:fitted_spectra}. For the CE component we fix $c=0$, as was suggested in \cite{Martinelli2022}. The $a$, $b$ and $c$ we call the spectral parameters. In Equation \eqref{eq:amplitude} we use the number of electrons and positrons $N_{\slice}$ in the slice. This effectively removes the scaling of the signal with the primary energy, as we see in Figure \ref{fig:spectral_b} where the values for all primary energies line up nicely.

In Figure \ref{fig:spectral_b} we plot the value of the GEO spectral $b$ parameter against the \textXmax\ of the shower it was extracted from, in three different slices. We do the same for all other spectral parameters. Guided by the observation that for each parameter all the points line up irrespective of primary type or energy, we fit a parabola to them, resulting in the spectral functions. These depend explicitly on \textXmax\ and implicitly on the selected slice and antenna as well as the geometry. We obtain a spectral function for every parameter, emission component, slice and antenna. They inform us about how the spectral parameters depend on the \textXmax\ of the shower in every slice and antenna, for this geometry.

To synthesise the emission from a shower with the target profile, we rescale that of the origin shower in every slice and antenna separately. For this we use the $\Xmax^{\text{origin}}$ of the origin shower and $\Xmax^{\text{target}}$ of the target profile to evaluate the spectral functions. From these we obtain the spectral parameters in every slice, with which we calculate the amplitude spectra, using Equation \eqref{eq:amplitude}. Through the ratio of these, we rescale the emission from the origin shower. 

As an example, if we were using the blue dashed shower in Figure \ref{fig:fitted_spectra} to synthesise the emission from the blue dotted shower, template synthesis would multiply the spectrum from the origin shower (the blue dashed line) by the ratio of the corresponding blue solid lines. Applying this procedure in every slice and then summing all the slices, we obtain the synthesised signal in the antenna.

The phase frequency spectra are not explicitly considered in templates synthesis. We assume these depend on the viewing geometry, which in our method is kept the same throughout the synthesis process. As such we take them to be a constant of the slice under consideration and simply take the phase spectrum from the origin shower to synthesise the target signal.

\section{Synthesis for inclined geometries}

\begin{figure}
	\begin{subfigure}[b]{0.5\textwidth}
		\centering
		\includegraphics[width=0.98\textwidth]{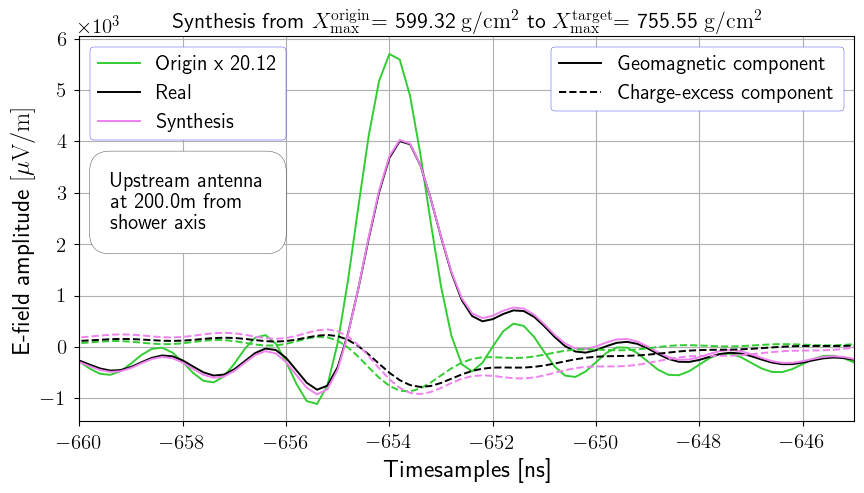}
		\caption{Early antenna at 200m from the shower axis in the shower plane, on the negative $\vec{v} \times \left( \vec{v} \times \vec{B} \right)$ axis.}
	\end{subfigure}
	~
	\begin{subfigure}[b]{0.5\textwidth}
		\centering
		\includegraphics[width=0.98\textwidth]{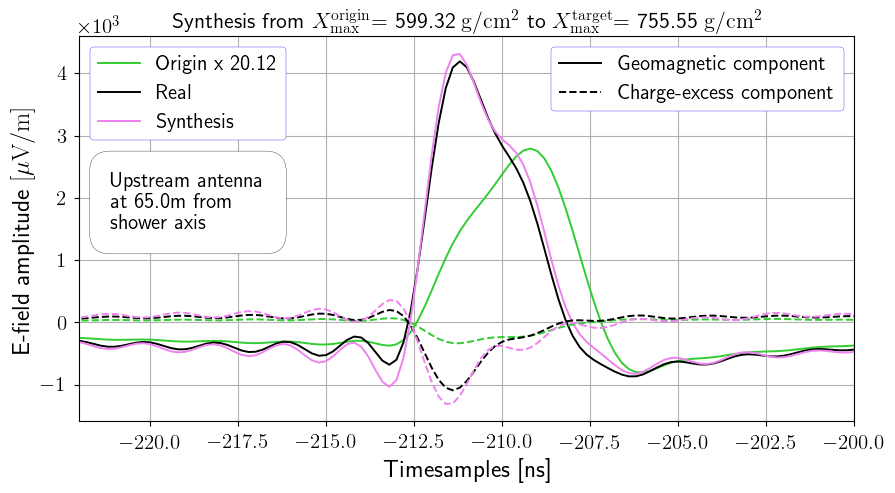}
		\caption{Early antenna at 65m from the shower axis in the shower plane, on the negative $\vec{v} \times \left( \vec{v} \times \vec{B} \right)$ axis.}
	\end{subfigure} \\
	\begin{subfigure}[b]{0.5\textwidth}
		\centering
		\includegraphics[width=0.98\textwidth]{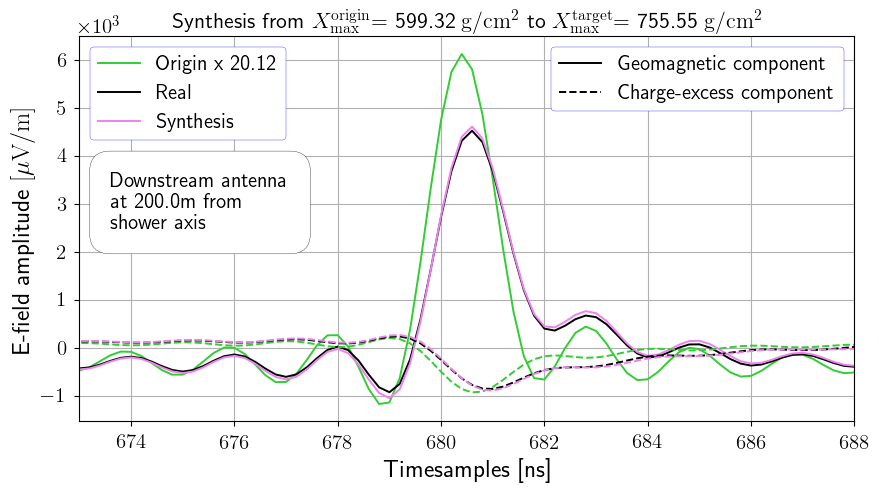}
		\caption{Late antenna at 200m from the shower axis in the shower plane, on the positive $\vec{v} \times \left( \vec{v} \times \vec{B} \right)$ axis.}
	\end{subfigure}
	~
	\begin{subfigure}[b]{0.5\textwidth}
		\centering
		\includegraphics[width=0.98\textwidth]{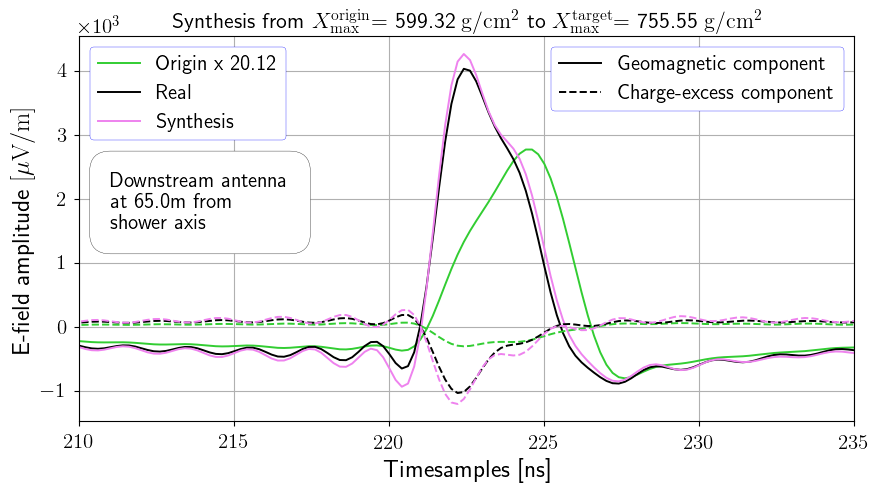}
		\caption{Late antenna at 65m from the shower axis in the shower plane, on the positive $\vec{v} \times \left( \vec{v} \times \vec{B} \right)$ axis.}
	\end{subfigure}
	\caption{Examples of synthesised pulses, for inclined showers coming from the South under a zenith angle $\theta=45 \degree$. We multiply the components of the origin shower with the ratio of primary energies to show the shift in pulse shape more clearly. Here we also simulated the target shower using CORSIKA and CoREAS, in order to be able to compare the result of template synthesis with it. The bottom plot shows the signals in one antenna, split into the GEO (solid lines) and CE (dashed lines) components. In green and black the signals as calculated by CoREAS are shown. Overlayed in magenta we show the synthesised pulse. The antennas at 200m are outside of the Cherenkov ring in the shower plane, while those at 60m are sitting inside of it and hence have a wider pulse shape.}
	\label{fig:synthesis}
\end{figure}

Before we can apply template synthesis to a shower with given geometry, we need the spectral functions for that geometry as well as an origin MC simulation. To obtain the former we generate a dedicated set of MC simulations with the geometry of interest. In the future we will investigate how these functions depend on the geometry, such that we are able to calculate them without the need of dedicated simulation set.

For the vertical geometry we performed an analysis to test the performance of template synthesis \cite{Desmet2023}. We compared the result of method to the signal coming from CoREAS, by taking the ratio of the peak amplitudes and energy fluences. We found the synthesis quality to depend strongly on $\Delta X_{\text{max}} = \Xmax^{\text{target}} - \Xmax^{\text{origin}}$. This can be interpreted as a measure of how far template synthesis has to extrapolate the origin signal, hence it is to be expected that it performs best when the \textXmax\ of origin and target are similar.
Still, the ratios stay within 10\% across the range
\begin{align*}
	-200 \gcm < \Delta X_{\text{max}} < 150 \gcm \; ,
\end{align*}
for all antennas we considered. Furthermore, deviations are always smaller than the scatter arising due to limitations in air shower universality, seeing that shower universality is intrinsically not fulfilled beyond the 10\% level. Hence it is not a limitation of the method, but rather shows there are other effects we did not account for. Investigating what those are and how they can be incorporated in template synthesis will be the subject of future work. 

\subsection{Applying template synthesis to an inclined geometry}

In Figure \ref{fig:synthesis} we show synthesised pulses for an inclined geometry, namely one with a zenith angle $\theta = 45 \degree$ coming from the South, i.e. the azimuth angle $\phi = 0 \degree$ in the CORSIKA coordinate system. For this geometry we simulated 200 showers, half of which had a proton primary with the other half having an iron primary. The primary energies were log-uniformly distributed between $10^{17} \eV$ and $10^{19} \eV$. When comparing the origin and target pulses, we see that they differ much more significantly in pulse shape than the vertical case. Still, template synthesis is able to handle this effect.

We show the pulses for 4 different antennas across the $\vec{v} \times \left( \vec{v} \times \vec{B} \right)$ axis. We chose two lateral distances in the shower plane, one inside and one outside the Cherenkove cone, and performed the synthesis in both the early and late antenna. As we are now considering an inclined geometry, the circular symmetry around the shower axis is broken. Still, template synthesis can handle both sides without issues.

When comparing the results from template synthesis in the vertical case \cite{Desmet2023} to those from the inclined geometry, it does seem like the method has more difficulties with the latter. There are several possible causes for this. First and foremost, the spectral functions were extracted from much smaller simulation set for the inclined case. This increases the probability that the spectral functions are influenced by outliers. Second, atmospheric effects become much more pronounced for more inclined air showers. Specifically the difference in density between the very early and very late slices is something we plan on investigating further. Lastly, also the evolution of the refractive index within a slice might play a role, although this is not expected to be a large effect.

\subsection{Using template synthesis for analyses}

The plots in Figure \ref{fig:synthesis} took a couple of minutes to produce, most of which was spent on the actual synthesis process. Even though the code has not been optimised, this is already orders of magnitude smaller than a typical MC simulation, which can have runtimes on the order of days at these energies.

Crucially, the scaling relations for this geometry were extracted beforehand. It is clear that we will not want to perform this operation for every geometry we want to consider. The goal will thus be to understand how the scaling relations depend on the air shower geometry. This will allow us to calculate the scaling relations for any desired geometry. With these in hand, we only need one or more origin showers in order to synthesise the emission for any shower with the specified geometry. 

Considering the ranges in which the method performs optimally mentioned above, no more than five origin showers would be necessary to cover the \textXmax\ range of interest. If we pick the origin showers to have their \textXmax\ in intervals of about 200 \textgcm , there will be one within the optimal interval for every shower profile we wish to synthesise. As these origin showers are the only MC simulations needed, we can effectively reduce the computation time for an analysis by a factor 10. 

Here the power of template synthesis really becomes clear. This also opens up the avenue of R/L reconstruction from data, an analysis which requires about a hundred simulations \cite{Corstanje2022}.

\subsection{Scaling relations in template synthesis}

In its current implementation, template synthesis only uses the longitudinal profile to characterise the target signal. From the profile, the number of emitting particles as well as the shower age can be determined for every slice. By rescaling with the particle number we also take care of any influences due to the primary type and energy.

However, it is clear that other parameters influence the radio signal as well. Some of these were already identified in the context of Radio Morphing \cite{Zilles2020}, such as the air density and index of refraction in the slice. The latter affects the coherency of the signal, which can be captured by the local Cherenkov angle. The former is more complicated and its effects are not fully understood. Several parametrisations have been performed, see for example \cite{Glaser2016, Chiche2021}. We are currently investigating both of these in the context of template synthesis.

The aforementioned effects can already impact synthesis when working with a fixed geometry. However, we would also like to generalise the method across geometries. For this we will need to understand how the spectral functions depend on the geometry. But when changing geometries, other aspects will have to be taken into account as well. An obvious one is the distance $R$ between the slice (i.e. the emission point) and the antenna. This is easily handled by scaling with $1 / R$. Next to this, we also need to take care of the change in geomagnetic angle, which should usually be straightforward as well.
	
\section{Conclusion and outlook}

We have developed a novel, hybrid method to synthesise the radio emission from extensive air showers. It makes use of a parametrised function in order to characterise the amplitude frequency spectrum of the emission at different points during the shower evolution. The parameters of this function can be related to the shower age through the spectral functions, by using the \textXmax\ in an atmospheric slice as a proxy for it. With the spectral functions the emission from a single MC simulation can be rescaled to synthesise the emission from an air shower with same geometry, but an arbitrary longitudinal profile. In previous work we showed this method performs well for vertical air showers. A detailed analysis for the vertical geometry shows we are limited by our understanding of shower-to-shower fluctuations.

Here we showed our first results for inclined air showers, which, at first glance, yielded good results. We compared early to late antennas, with the same lateral distance in the shower plane, and found no significant discrepancy between the two cases. This shows that template synthesis is capable of handling the inherent asymmetries associated with non-vertical geometries.

For our next steps, we will analyse other inclined geometries as well, to investigate whether there is a limit in zenith angle that our method can handle. Once we reach the zenith angles where Radio Morphing becomes applicable, we plan to compare the results of template synthesis to it. 

An arguably more important step however, is to understand the scaling with geometry and atmosphere. In the current version of template synthesis, the spectral functions are determined for every slice, antenna and geometry. Future work will investigate how these functions can be generalised across slices and geometries, as to be able to apply the method to obtain the emission from any desired EAS.

\bibliographystyle{JHEP}
\bibliography{sources}

\end{document}